\documentclass{basi}
\usepackage{graphicx}
\usepackage{rotating}
\begin{document}
\title[C\,{\sc{ii}} column densities along Galactic sight-lines]
      {Estimation of C\,{\sc{ii}} and C\,{\sc{ii}$^\ast$} column densities along Galactic sight-lines} 
\author[V.~S. Parvathi et al.]{Veena S. Parvathi,$^{1}$\thanks{E-mail: veena$\_$sivaram@yahoo.co.in (VSP), 
                               sofia@american.edu (UJS), jmurthy@yahoo.com (JM), brsbabu@gmail.com (BRSB)} 
Ulysses J. Sofia,$^{2}$ Jayant Murthy$^{3}$ and B. R. S. Babu$^{1}$ \\
$^{1}$Department of Physics, University of Calicut, Kerala 673635, India \\
$^{2}$Department of Physics, American University, 4400 Massachusetts Avenue, NW, \\ Washington, DC 20016, USA \\ 
$^{3}$Indian Institute of Astrophysics, II Block, Koramangala, Bangalore 560034, India }
\date{Received 2010 September 15; accepted 2010 October 19}

\maketitle

\label{firstpage}

\begin{abstract}
We present interstellar C\,{\sc{ii}} (1334.5323 \AA) and C\,{\sc{ii}$^\ast$} (1335.7077 \AA) column density measurements along 14 Galactic 
sight-lines.  These sight-lines sample a variety of Galactic disk environments and include paths that range nearly two orders of magnitude in average hydrogen densities ($<$n(H)$>$) along the lines of sight.  Five of the sight-lines show super-Solar gas phase abundances of carbon.  Our results show that the excess carbon along these sight-lines may result from different mechanisms taking place in the regions associated with these stars. 
\end{abstract}


\section{Introduction}
\label{sec:intro}
The interstellar medium (ISM) is a mixture of dust and gas and is responsible for interstellar extinction of starlight.  Extinction is
produced by both absorption and scattering and is generally represented by a plot of A$_{\lambda}$/Av (wavelength dependent
extinction/extinction at V magnitude) against $\lambda^{-1}$ (inverse of wavelength).  The dominant features in the extinction curve are the
bump centered at 2175 \AA ~ and the non-linear Far Ultraviolet (FUV) rise (Fitzpatrick \& Massa, 1986, 1988; hereafter referred to as FM).
The strength of the bump and the curvature of the FUV rise varies from one sight-line to another.  The bump in the extinction curves of the
Galactic sight-lines is very strong and must be produced by an abundant element in the ISM.  Since sulphur is believed to be at most weakly
depleted in the diffuse ISM (Colangeli et al. 2003), and since nitrogen and oxygen are electron acceptors, the carrier of the bump must
contain one or more elements from the four member set {C, Mg, Si, Fe} (Draine 1989).  Clayton \& Martin (1985) noted that the sequence Small
Magallanic Cloud (SMC), Large Magallanic Cloud (LMC) and Milkyway (MW) appear to have an increasing strength of the 2175 \AA ~ bump and an
increasing C/O ratio, suggesting that the carrier of the 2175 \AA ~ bump feature may be carbonaceous.  A majority of the proposed models for
the 2175 \AA ~ feature are based on carbon: Polycyclic Aromatic Hydrocarbon (PAH; Joblin, Leger, \& Martin 1992), graphite (Draine 1989), amorphous carbon (Duley \& Williams 1981), Quenched Carbonaceous Chondrites (QCCs; Sakata et al. 1977) and Hydrogenated Amorphous Carbon (HAC; Mennella \& Colangeli 1999).  

\begin{figure}
\centering
\includegraphics[width=13.5cm,height=10cm]{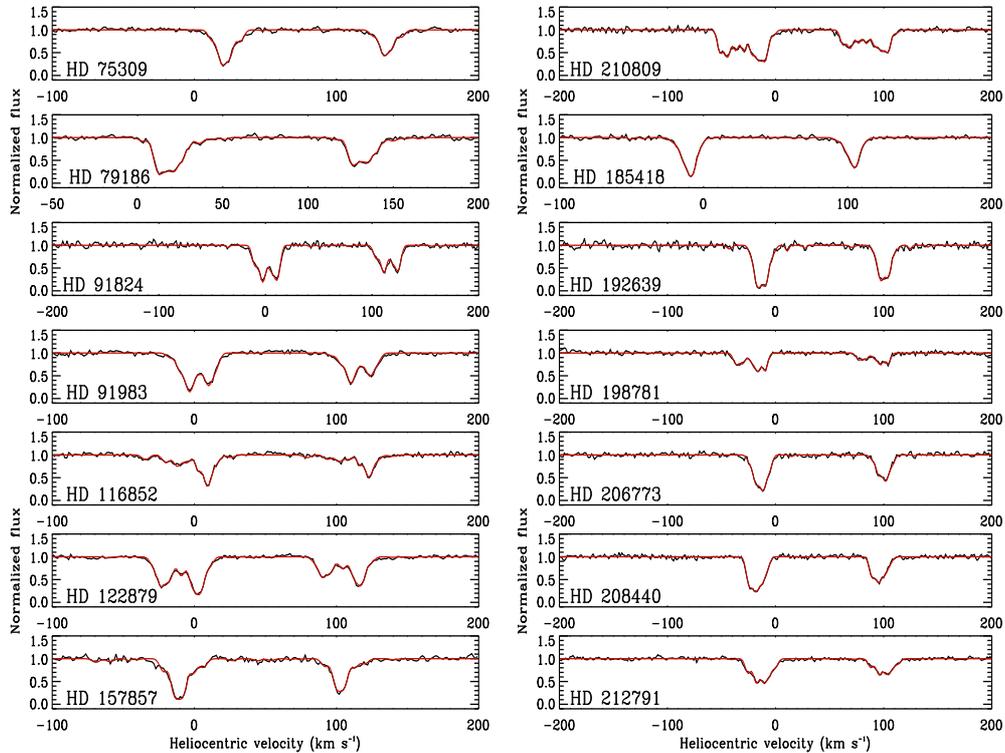}
\caption{Profile fits to the $\lambda\lambda$ 1239, 1240 \AA ~ transitions of Mg\,{\sc{ii}} for the fourteen lines of sight.  Dark lines represent normalized flux and red lines represent the model fits.}
\label{figure:MgII_fits}
\end{figure}

Carbon is an important element in the ISM and is the fourth most abundant element after hydrogen, helium and oxygen.  Carbon is the dominant heat source for the diffuse ISM.  Carbon grains heat the interstellar gas through photoelectric mechanism and recombination of electrons with the dust grains (Bakes \& Tielens 1994).  Interstellar cloud cooling is through the rotational and fine structure lines CO and C$^{+}$ (Bakes \& Tielens 1994).  In order to fully understand the ISM, it is important to know the distribution of carbon among its different phases.   

C\,{\sc{ii}} is the dominant form of carbon in neutral interstellar clouds (clouds containing H\,{\sc{i}} and H$_{2}$) because of its
ionization potential (11.26 eV).  Despite its importance, there are only very few abundance measurements of interstellar carbon (Hobbs, York
\& Oegerle 1982; Cardelli et al. 1993; Cardelli et al. 1996; Sofia et al. 1997; Sofia, Fitzpatrick \& Meyer 1998; Sofia et al. 2004).  This
may be due to the difficulties associated with the measurement of C\,{\sc{ii}} column density as the ion produces only very strong and very
weak absorption features in typical neutral sight-lines.  The strong C\,{\sc{ii}} transitions at 1036.3367 \AA ~ and 1334.5323 \AA ~ are
highly saturated and the column density measurements from these transitions are associated with large uncertainties.  The transition at
2325.4029 \AA ~ is too weak to be detected for many of the sight-lines.  The column density measurements mentioned in the above works are all from the weak transition at 2325.4029 \AA. 
 
C\,{\sc{ii}} transitions are always associated with excited state (C\,{\sc{ii}$^\ast$}) transitions and more often with 
$^{13}$C\,{\sc{ii}} and its excited state ($^{13}$C\,{\sc{ii}$^\ast$}) transitions.  Sofia \& Parvathi (2009) (hereafter referred to as Paper
1) have developed a method to accurately determine the C\,{\sc{ii}}  and C\,{\sc{ii}$^\ast$} column densities by considering the dominant
transitions at 1334.5323 \AA ~ and 1335.7077 \AA ~ together with their nearby transitions at 1334.519 \AA ~ ($^{13}$C\,{\sc{ii}}), 
1335.649 \AA ~ ($^{13}$C\,{\sc{ii}$^\ast$}), 1335.6627 \AA ~ ($^{12}$C\,{\sc{ii}$^\ast$}) and 1335.692 \AA$ $ ($^{13}$C\,{\sc{ii}$^\ast$}).   
In Paper 1, 
they have used this method to determine the interstellar C\,{\sc{ii}} and C\,{\sc{ii}$^\ast$} column densities along six translucent sight-lines 
(sight-lines having A$_{V}$ $>$ 1).  In the present work, we have extended this method to 14 sight-lines observed by the Space Telescope 
Imaging Spectrograph (STIS) aboard the Hubble Space Telescope (HST). 

\section{Methodology}
\begin{table}[b]\small
\caption{List of UVBLUE models used} 
\begin{center}
\begin{tabular}{l c l c}
\hline
Target star & UVBLUE model & Target star & UVBLUE model \\ [0.5ex]
\hline
HD 75309 & t28000g40p00k2.flx.bz2 & HD 185418 & t13000g40p00k2.flx.bz2 \\
HD 79186 & t14000g40p00k2.flx.bz2 & HD 192639 & t19000g40p00k2.flx.bz2  \\
HD 91824 & t22000g40p00k2.flx.bz2 & HD 198781 & t22000g40p00k2.flx.bz2  \\
HD 91983 & t20000g40p00k2.flx.bz2 & HD 206773 & t15000g40p00k2.flx.bz2  \\
HD 116852 & t13000g40p00k2.flx.bz2 & HD 208440 & t25000g40p00k2.flx.bz2  \\
HD 122879 & t09000g40p00k2.flx.bz2 & HD 210809 & t30000g40p00k2.flx.bz2  \\
HD 157857 & t20000g40p00k2.flx.bz2 & HD 212791 & t18000g40p00k2.flx.bz2  \\ [1ex]
\hline
\end{tabular}
\end{center}
\label{table:UVBLUE}
\end{table}

The 14 sight-lines chosen in the present study probe a variety of Galactic disk environments and include paths that range nearly two orders of magnitude in average sight-line hydrogen density ($<$n(H)$>$). Amongst all the sight-line parameters, $<$n(H)$>$ is the one that shows the strongest correlation with the elemental abundances for a given set of sight-lines.  Therefore the depletion along a given line of sight can be accurately represented as a function of $<$n(H)$>$.  The sight-lines in our sample were earlier studied for abundances of 
Mg\,{\sc{ii}}, P\,{\sc{ii}}, Mn\,{\sc{ii}}, Ni\,{\sc{ii}}, Cu\,{\sc{ii}} and Ge\,{\sc{ii}} by Cartledge et al. (2006).  The high resolution STIS
observations used in the present study are all taken from the archive at the Space Telescope Science Institute (STScI).  STIS datasets have
the advantage that they contain other spectral features required for our analysis (e.g. transitions of Mg\,{\sc{ii}} at 1239 and 1240 \AA).
In these data, each exposure was made using either of the two STIS setups; the 0.2$^{\prime\prime}\times$0.2$^{\prime\prime}$ aperture with
the E140H grating centered at 1271 \AA$ $ or the same grating exposed through the 0.2$^{\prime\prime}\times$0.09$^{\prime\prime}$ aperture.
These data have never been used for accurately determining interstellar carbon column densities by the method of profile fitting.  This is
because of the intrinsic difficulties in deriving accurate column density values from the damping wings of the C\,{\sc{ii}} transition.  Inaccurate separation of the Lorentzian damping wings from the continuum and lack of any cloud (absorbing component) structure information along the sight-line can lead to large uncertainties in the result.

\begin{figure}
\centering
\includegraphics[width=13.5cm,height=10cm]{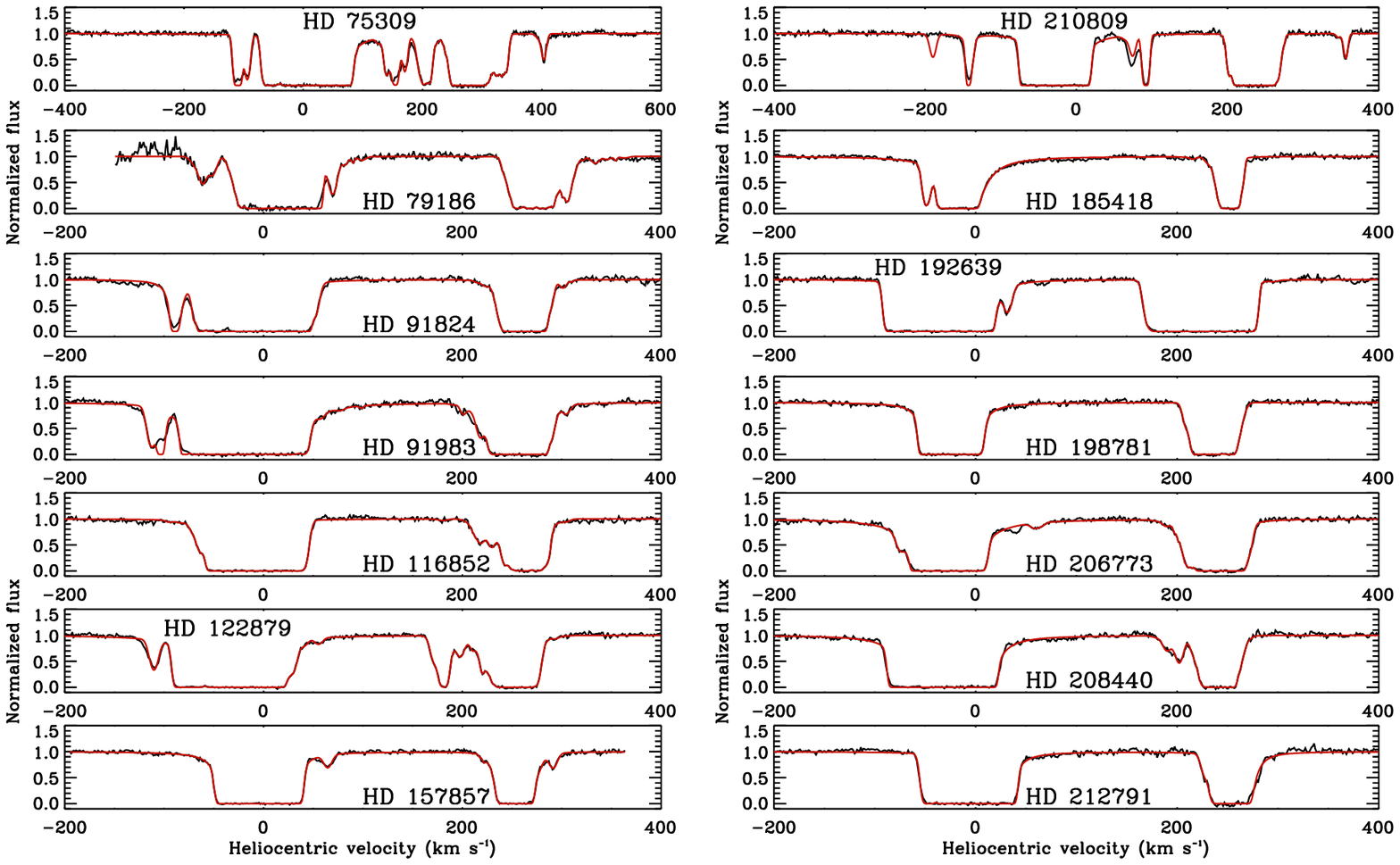}
\caption{Profile fits to the C\,{\sc{ii}} and C\,{\sc{ii}$^\ast$} transitions together with their associated $^{13}$C transitions.  As the $^{13}$C and $^{12}$C transitions are blended, they are difficult to distinguish from each other.  Dark lines represent normalized flux and red lines represent model fits.}
\label{figure:CII_fits}
\end{figure}

\begin{sidewaystable} \small
\renewcommand{\thefootnote}{\alph{footnote}}
\caption{Column density measurements of C\,{\sc{ii}} and C\,{\sc{ii}$^\ast$}}
\begin{center}
\begin{tabular}{l c c c c c c c}
\hline
	    & \multicolumn{2}{c}{Galactic co-ord.}	\\
\cline{2-3}
Target star & l & b & log$_{10}$[N(H)]\footnotemark & log$_{10}<$n(H)$>$\footnotemark & N(C\,{\sc{ii}}) & N(C\,{\sc{ii}}$^\ast$) 
                                                                                                             & 12+log$_{10}$[N(C)/N(H)]\\
	   &               &                 &                   &                   & ($\times$10$^{17}$                  &                  \\
	   &               &                 & (atoms cm$^{-2}$) & (atoms cm$^{-3}$) &      atoms cm$^{-2}$) & (atoms cm$^{-2}$)\\
\hline
HD 75309$^{\star}$ & 265.85 & $-$01.89 & 21.18(0.08) & $-$0.55  & 13.3$^{0.1}_{0.1}$ & 1.44e+17 & 8.99$^{0.08}_{0.08}$ \\[0.5ex]
HD 79186 & 267.36 & +02.25 & 21.42(0.07) & $-$0.31  & 0.15$^{0.78}_{0.01}$ & 1.26e+15  & 6.80$^{0.79}_{0.07}$ \\[0.5ex]
HD 91824$^{\star}$ & 285.69 & +00.06 & 21.16(0.05) & $-$0.73  & 6.89$^{0.03}_{0.05}$ & 3.53e+16 & 8.70$^{0.05}_{0.05}$ \\[0.5ex]
HD 91983$^{\star}$ & 285.87 & +00.05 & 21.24(0.08) & $-$0.65  & 7.89$^{0.001}_{0.09}$ & 5.06e+16 & 8.68$^{0.08}_{0.08}$ \\[0.5ex]
HD 116852 & 304.88 & $-$16.13 & 21.02(0.08) & $-$1.15  & 2.61$^{0.04}_{0.05}$ & 9.23e+14 & 8.40$^{0.08}_{0.08}$ \\[0.5ex]
HD 122879 & 312.26 & +01.79 & 21.34(0.10) & $-$0.47 & 6.70$^{0.04}_{0.04}$ & 5.87e+15 & 8.49$^{0.10}_{0.10}$ \\[0.5ex]
HD 157857 & 012.97 & +13.31 & 21.44(0.07) & $-$0.41 &  4.89$^{0.18}_{0.20}$ & 7.75e+16 & 8.31$^{0.07}_{0.07}$ \\[0.5ex]
HD 185418 & 053.60 & $-$02.17 & 21.41(0.07) & +0.08 &  6.43$^{0.19}_{0.29}$ & 8.18e+15 & 8.40$^{0.07}_{0.07}$ \\[0.5ex]
HD 192639 & 074.90 & +01.47 & 21.48(0.07) & $-$0.27 & 3.46$^{0.07}_{0.08}$ & 3.28e+16 & 8.10$^{0.07}_{0.07}$ \\[0.5ex]
HD 198781 & 099.94 & +12.61 & 21.15(0.06) & $-$0.13 & 3.36$^{0.24}_{0.21}$ & 2.41e+15 & 8.38$^{0.07}_{0.07}$ \\[0.5ex]
HD 206773$^{\star}$ & 099.80 & +03.61 & 21.25(0.05) & $-$0.03 & 12.6$^{0.002}_{1.0}$ & 8.93e+16 & 8.88$^{0.05}_{0.06}$ \\[0.5ex]
HD 208440$^{\star}$ & 104.02 & +06.43 & 21.32(0.08) & +0.04 & 8.10$^{0.16}_{0.20}$ & 4.31e+15 & 8.59$^{0.08}_{0.08}$ \\[0.5ex]
HD 210809 & 099.84 & $-$03.12 & 21.33(0.08) & $-$0.70  & 6.18$^{0.10}_{0.18}$ & 5.76e+16& 8.50$^{0.08}_{0.08}$  \\[0.5ex]
HD 212791 & 101.64 & $-$04.30 & 21.22(0.09) & +0.16 & 3.36$^{0.09}_{0.03}$ & 1.45e+17 & 8.46$^{0.09}_{0.09}$ \\[0.5ex]
\hline
\end{tabular}
\end{center}
\label{table:carbon}
\footnotesize{The sight-lines marked with $\star$ show super-Solar abundance of carbon.} \\
{$^a$}\footnotesize{log$_{10}$[N(H)] values are obtained from Cartledge et al. (2006).} \\
{$^b$}\footnotesize{log$_{10}<$n(H)$>$ values are obtained from Cartledge et al. (2006).}
\renewcommand{\thefootnote}{\arabic{footnote}}
\end{sidewaystable}

\addtocounter{footnote}{-2}
We have accurately separated the damping wings from the continuum by using existing stellar models to fit the stellar continuum in the data.  Specifically, we employed UVBLUE\footnote{http://www.inaoep.mx/$\sim$modelos/uvblue/uvblue.html}, a high-resolution library of synthetic spectra of stars covering the ultraviolet wavelength range, developed by the Instituto Nacional de Astrofisica Optica y Electronica (INAOE) of Pubela (Mexico), the Brera Astronomical Observatory of Milan (Italy) and the Bologna Astronomical Observatory (Italy).  The stellar models covering a large range of temperatures, surface gravities and metallicities are computed with the Kurucz (2003) set of model atmospheres, by means of the ATLAS9/SYNTHE code (Kurucz 2005).  Information from the SIMBAD\footnote{http://simbad.u-strasbg.fr/simbad/} database is used to identify the target star's spectral type and a scan over the parameter space about that value is made to find the appropriate stellar model.  In addition to that, we have modified the stellar model to account for stellar reddening, difference in metallicity from the model, rotational velocity of the star and the star's radial velocity shift\footnote{Steps followed to modify the UVBLUE model are described in detail in Paper 1}.  We thereby obtained a model which can well represent the stellar continuum in the data.  As the spectrum of HD 91824 has very complex structure for its 1334.5323 \AA$ $ transition, we normalized the continuum by supplementing the UVBLUE fit with a fifth order polynomial.  Details of the UVBLUE models used for different lines of sight are given in Table \ref{table:UVBLUE}.

\begin{figure}[t]
\centering
\includegraphics[width=13cm,height=8cm]{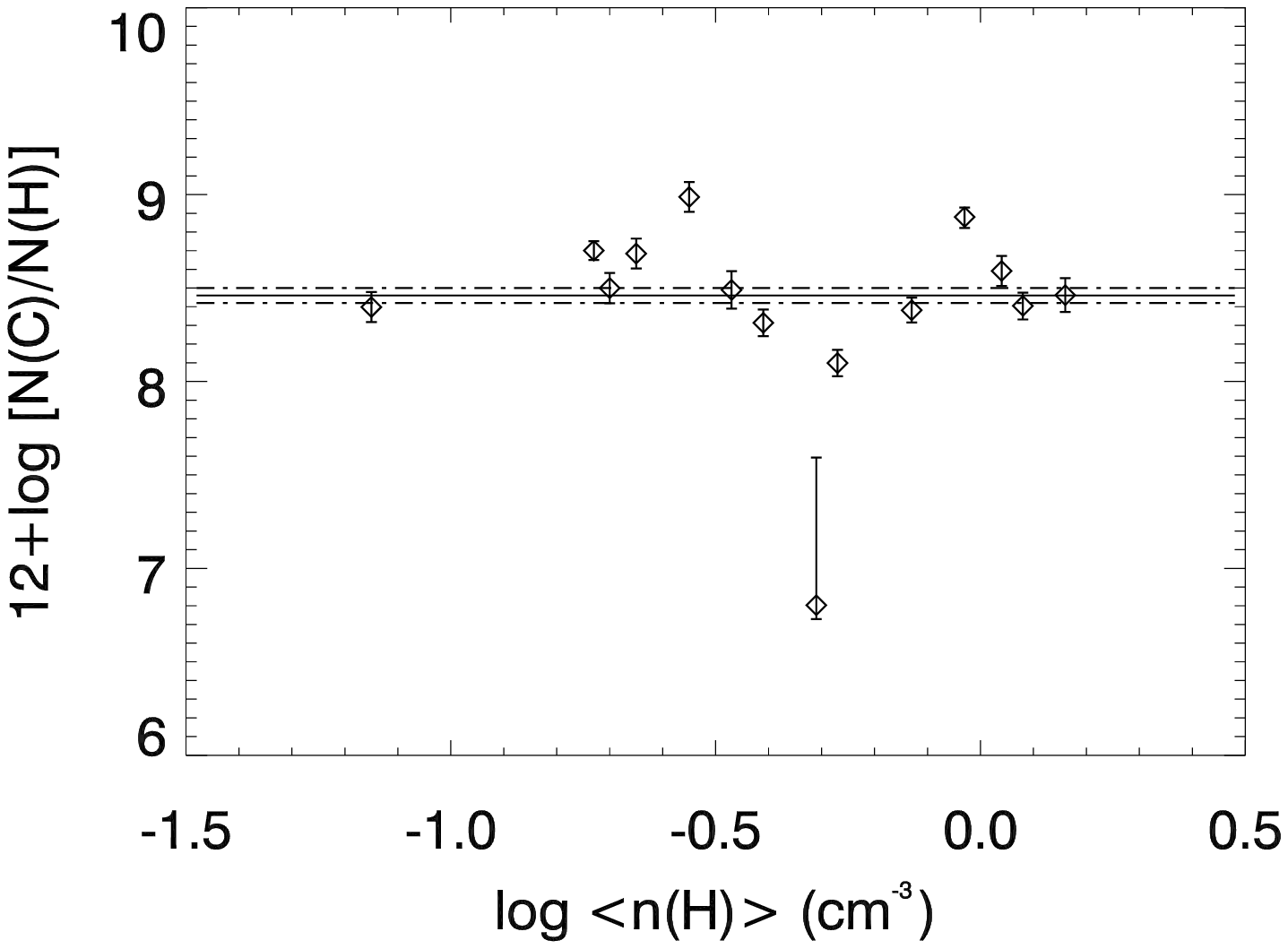} 
\caption{Gas-phase abundance of carbon on a logarithmic scale plotted against the logarithm of average hydrogen density.  Associated error bars for the data points can also be seen in the figure.  The solid straight line with an abundance value of 8.46 corresponds to the Solar system abundance of carbon (Lodders 2003) and the dashed lines represent associated 1$\sigma$ error bars.}
\label{figure:ACvsn_H}
\end{figure}
  
The cloud structure information for the sight-lines were obtained by fitting the Mg\,{\sc{ii}} transitions at $\lambda \lambda$ 1239, 1240
\AA. ~ The Mg\,{\sc{ii}} profile fits for the fourteen lines of sight are shown in Figure \ref{figure:MgII_fits}.  We have constructed the
absorption line profile using the parameters heliocentric velocities of the clouds (v$_{\odot}$), Doppler broadening parameters of the
clouds (b), column densities of the species in the clouds (N) and the shift in the velocity vector due to the instrumental uncertainty
(v$_{shift}$).  The atomic constants of the transition are obtained from Morton (2003).  In order to find the set of parameter values that
produces the best fit, we have employed a standard IDL routine
`MPFITFUN'\footnote{http:$//$cow.physics.wisc.edu/$\sim$craigm$/$idl$/$mpfittut.html} (Markwardt 2009), which performs Levenberg-Marquardt
least-squares fit to an IDL function.  MPFITFUN fits a user-supplied model in the form of an IDL function to a set of user-supplied data.
Given the data and their uncertainties, MPFITFUN finds the best set of model parameters which match the data and returns them in an array.
We have the component structure information from Mg\,{\sc{ii}} transitions.  Once again we have made use of MPFITFUN to fit the 
C\,{\sc{ii}} and C\,{\sc{ii}$^\ast$} transitions, with an additional model parameter `N$_{star}$'.  This parameter N$_{star}$ signifies 
the fraction of C\,{\sc{ii}} existing in the form of C\,{\sc{ii}$^\ast$}.  We have also used the terrestrial value for the isotope 
ratio of $^{12}$C to $^{13}$C.  The C\,{\sc{ii}} and C\,{\sc{ii}$^\ast$}  profile fits for the fourteen lines of sight are shown 
in Figure \ref{figure:CII_fits}.     

\section{Results and Discussion}

The C\,{\sc{ii}} and C\,{\sc{ii}$^\ast$}  column density measurements for the 14 sight-lines in our sample are listed in Table
\ref{table:carbon}.  We have also calculated the gas-phase carbon abundances along the sight-lines by adopting Lodders' Solar system value
of 12+log$_{10}$(C/H) = 8.46$\pm$0.04 (Lodders 2003).  It is assumed that interstellar elemental abundances can be well substituted by Solar
abundances.  Figure \ref{figure:ACvsn_H} shows gas-phase carbon abundance plotted against logarithm of average hydrogen density.  Five of
the sight-lines show super-Solar abundance of carbon in gas-phase.  The deviation in the positive direction from the Solar value is large
for the sight-lines toward HD 75309 and HD 206773 whereas it can be treated as marginal for the sight-lines toward HD 91824, HD 91983
and HD 208440.  HD 75309 is located in the Vela supernova remnant (Vela SNR) (Nichols \& Slavin 2004).  The super-Solar carbon abundance
along this sight-line can be explained by dust in this evolving SNR being first heated and then destroyed by shock wave (Vancura et al.
1994).  The dust destruction leads to the formation of gas. HD 206773 is a member of Cepheus OB2 association (Pan et al. 2005).  Based on
measurements of CO emission, Patel et al. (1998) suggested that the bubble was created likely by stellar winds and supernova explosions.
From the 21 cm data of H\,{\sc{i}}, \'{A}brah\'{a}m, Bal\'{a}zs \& Kun (2000) proposed that a supernova explosion might have occurred as recently as about 2 Myr ago.  This explosion might have expanded the pre-existing bubble further.  The star in its final stages of evolution ejects its outer layers by a supernova explosion or planetary nebula.  Carbon is one among those heavy elements which are likely to be ejected during a supernova explosion.  The ejected elements get mixed up with the surrounding ISM thereby creating inhomogeneities in elemental abundances.  Therefore the different mechanisms taking place in the regions surrounding HD 75309 and HD 206773 would have favoured the formation of more gas-phase carbon along these 
sight-lines.  

The sight-line towards HD 79186 is interesting as it shows very large deviation from the Solar abundance unlike other sight-lines in our
sample.   This sight-line shows very low gas-phase abundance of carbon with very high error bar.  From Figure \ref{figure:CII_fits} it is
evident that the blue wings of the C\,{\sc{ii}} line is not continuum fitted very well.  Our analysis relies on continuum fits based on the
UVBLUE stellar spectral library.  Although the models explore a wide range in parameter space, they cannot be expected to always fit our
stellar continuua exactly.  HD 79186 is one of the cases where the continuum of the star differs from the UVBLUE models.  When the continuum
fits are less accurate, the uncertainties in the C\,{\sc{ii}} column densities are higher.  Because of the stochastic nature of the continuum 
mismatch, it is difficult to quantify this uncertainty in the reported errors.  However, to be consistent with our method, we avoided using additional low order polynomials to accurately fit the stellar continuum of HD 79186.  A very low gas-phase abundance of carbon along the sight-line towards HD 79186 implies the presence of more carbon containing dust along this sight-line.

It is also observed from Figure \ref{figure:ACvsn_H} that the gas-phase abundance of carbon does not follow any strict relation with $<$n(H)$>$, unlike gas-phase abundances of other elements studied by Cartledge et al. (2006) all of which following a Boltzmann function description of abundance variation.  This is an indication of carbon alone possessing some unique characteristics.  However, we require a large sample of data in order to completely understand carbon.   
 
\section*{Acknowledgements}
This research was supported by the Hubble Space Telescope Science Institute grant HST-AR-11775 to Whitman College and American University and by the grant from Council of Scientific and Industrial Research, India.  This work has made use of the SIMBAD astronomical database operated at CDS, Strasbourg, the SAO/NASA Astrophysics Data System (ADS) and the Multimission Archive at Space Telescope Science Institute (MAST).  A major part of this work has been carried out from Indian Institute of Astrophysics (IIA), Bangalore and the remaining part from University of Calicut, Kerala, India.

\end{document}